\documentstyle[twocolumn,aps,prl,array]{revtex}
\begin{document}
\draft
\title{``All versus Nothing'' Inseparability for Two Observers}
\author{Ad\'{a}n Cabello\thanks{Electronic address:
adan@cica.es}}
\address{Departamento de F\'{\i}sica Aplicada II,
Universidad de Sevilla, 41012 Sevilla, Spain}
\date{\today}
\maketitle
\begin{abstract}
A recent proof of Bell's theorem without inequalities [A. Cabello,
Phys. Rev. Lett. {\bf 86}, 1911 (2001)] is formulated as a
Greenberger-Horne-Zeilinger--type proof involving just two
observers. On one hand, this new approach allows us to derive an
experimentally testable Bell inequality which is violated by
quantum mechanics. On the other hand, it leads to a new
state-independent proof of the Kochen-Specker theorem and provides
a wider perspective on the relations between the major proofs of
no-hidden-variables.
\end{abstract}
\pacs{PACS numbers: 03.65.Ud,
03.65.Ta,
42.50.-p}
\narrowtext


Bell's theorem \cite{Bell64} refutes local theories based on
Einstein, Podolsky, and Rosen's (EPR's) ``elements of reality''
\cite{EPR35}. A recently introduced proof without inequalities
\cite{Cabello01} presents the same logical structure as that of
Hardy's proof \cite{Hardy93}, but exhibits a greater contradiction
between EPR local elements of reality and quantum mechanics. Here
a simpler version of the proof in \cite{Cabello01} will be
introduced. This new version parallels Mermin's reformulation
\cite{Mermin90a} of Greenberger, Horne, and Zeilinger's (GHZ's)
proof \cite{GHZ89} and, besides being simpler, it emphasizes the
fact that \cite{Cabello01} is also an ``all versus nothing''
\cite{Mermin90b} or GHZ-type proof of Bell's theorem, albeit with
only two observers. In addition, this new approach will allow us
to derive an inequality between correlation functions which is
violated by quantum mechanics. Moreover, this new version will
also constitute the basis for a new state-independent proof of the
Kochen-Specker (KS) theorem \cite{KS67}. The whole set of new
results provides a wider perspective on the relations between the
most relevant proofs of no-hidden-variables.


Consider four qubits labeled 1, 2, 3, 4, prepared in the state
\begin{equation}
\left|{\psi } \right\rangle _{1234} = {1 \over 2}
\left( {
\left| 0011 \right\rangle -
\left| 0110 \right\rangle -
\left| 1001 \right\rangle +
\left| 1100 \right\rangle
} \right),
\label{twosinglets}
\end{equation}
which, as can be easily checked, is the product of two singlet
states, $\left| {\psi ^-} \right\rangle _{13} \otimes
\left| {\psi ^-} \right\rangle _{24}$.

Let us suppose that qubits 1 and 2 fly apart from qubits 3 and 4,
and that an observer, Alice, performs measurements on qubits 1 and
2, while in a spacelike separated region a second observer, Bob,
performs measurements on qubits 3 and 4.


By using the following notation: $z_i = \sigma_{zi}$, $x_i =
\sigma_{xi}$, and $z_i x_j = \sigma_{zi} \otimes \sigma_{xj}$,
etc.; and introducing $(\cdot)$ to separate operators or operator
products that can be viewed as EPR local elements of reality, it
is easy to check that the state $\left| \psi \right\rangle$
satisfies
\begin{eqnarray}
z_1 & \cdot & z_3 \left|{\psi } \right\rangle=
-\left|{\psi } \right\rangle,
\label{uno} \\
z_2 & \cdot & z_4 \left|{\psi } \right\rangle=
-\left|{\psi } \right\rangle,
\label{dos} \\
x_1 & \cdot & x_3 \left|{\psi } \right\rangle=
-\left|{\psi } \right\rangle,
\label{tres} \\
x_2 & \cdot & x_4 \left|{\psi } \right\rangle=
-\left|{\psi } \right\rangle,
\label{cuatro} \\
z_1 z_2 & \cdot & z_3 \cdot z_4 \left|{\psi } \right\rangle=
\left|{\psi } \right\rangle,
\label{cinco} \\
x_1 x_2 & \cdot & x_3 \cdot x_4 \left|{\psi } \right\rangle=
\left|{\psi } \right\rangle,
\label{seis} \\
z_1 \cdot x_2 & \cdot & z_3 x_4 \left|{\psi } \right\rangle=
\left|{\psi } \right\rangle,
\label{siete} \\
x_1 \cdot z_2 & \cdot & x_3 z_4 \left|{\psi } \right\rangle=
\left|{\psi } \right\rangle,
\label {ocho} \\
z_1 z_2 \cdot x_1 x_2 & \cdot & z_3 x_4 \cdot x_3 z_4
\left|{\psi } \right\rangle= - \left|{\psi } \right\rangle.
\label{nueve}
\end{eqnarray}


According to EPR, if Alice (Bob) can predict with certainty and
without in any way disturbing Bob's (Alice's) qubits, the value of
a physical quantity of Bob's (Alice's) qubits, then there exists
an element of physical reality corresponding to this physical
quantity \cite{EPR35}. Eqs.~(\ref{uno})-(\ref{nueve}) contain only
local (Alice's or Bob's) operators and allow Alice to infer EPR
local elements of reality for Bob's observables $z_3$, $z_4$,
$x_3$, $x_4$, $z_3 x_4$, and $x_3 z_4$; they also allow Bob to
infer EPR local elements of reality for Alice's observables $z_1$,
$z_2$, $x_1$, $x_2$, $z_1 z_2$, and $x_1 x_2$. In addition,
Eqs.~(\ref{uno})--(\ref{nueve}) allow Alice and Bob to predict the
following relations between the values of the elements of reality:
\begin{eqnarray}
v(z_1) & & v(z_3) = -1,
\label{one} \\
v(z_2) & & v(z_4) = -1,
\label{two} \\
v(x_1) & & v(x_3) = -1,
\label{three} \\
v(x_2) & & v(x_4) = -1,
\label{four} \\
v(z_1 z_2) & & v(z_3) v(z_4) = 1,
\label{five} \\
v(x_1 x_2) & & v(x_3) v(x_4) = 1,
\label{six} \\
v(z_1) v(x_2) & & v(z_3 x_4) = 1,
\label{seven} \\
v(x_1) v(z_2) & & v(x_3 z_4) = 1,
\label{eight} \\
v(z_1 z_2) v(x_1 x_2) & & v(z_3 x_4) v(x_3 z_4) = -1.
\label{nine}
\end{eqnarray}

However, it is impossible to assign values, either $-1$ or $+1$,
that satisfy Eqs.~(\ref{one})--(\ref{nine}), because when we take
the product of Eqs.~(\ref{one})--(\ref{nine}) each value appears
twice in the left-hand side, while the right hand side is $-1$. We
therefore conclude that the predictions of quantum theory for a
single copy of the state $\left|{\psi } \right\rangle$ cannot be
reproduced with any local model based on EPR's criterion of
elements of reality.


The GHZ proof of Bell's theorem provided an ``all versus nothing''
refutation of EPR elements of reality but required three or more
spacelike separated observers. The proof presented here, which is
an extension of \cite{Cabello01}, is an ``all versus nothing''
refutation which needs only two observers.

In an ideal situation, the contradiction with EPR elements of
reality would appear after many runs of nine different
experiments, one for each of Eqs.~(\ref{uno})--(\ref{nueve}).
These runs would accumulate evidence that the appropriate
correlations are strong enough to support elements of reality. By
using the results of eight of these experiments, one can make a
deduction about the results of the ninth experiment based on
elements of reality. According to quantum mechanics, such a
deduction would then be contradicted in every single run of this
experiment.

However, the same conclusion cannot be inferred directly from the
actual data in a nonideal laboratory realization of the experiment
because, for example, the efficiency of real detectors does not
allow us to observe the strong correlations assumed both in the
EPR original argument \cite{EPR35} and in the gedanken proofs of
Bell's theorem. To circumvent this problem it is common practice
to derive inequalities between experimentally observable
correlation functions whose validity relies on very general
probabilistic locality conditions but which are violated by the
corresponding quantum predictions \cite{CHSH69}. Next, I will
derive a Bell inequality for the state $\left|{\psi }
\right\rangle$ based on the previously introduced gedanken proof.
Such a derivation parallels Mermin's derivation of an inequality
for $n$ qubits in a GHZ state \cite{Mermin90b}.


All the relevant features of the gedanken proof follow from the
fact that $\left|{\psi } \right\rangle$ is an eigenstate of the
operator
\begin{eqnarray}
O & = & -z_1 \cdot z_3
-z_2 \cdot z_4
-x_1 \cdot x_3
-x_2 \cdot x_4 \nonumber \\
 & & +z_1 z_2 \cdot z_3 \cdot z_4
+x_1 x_2 \cdot x_3 \cdot x_4 \nonumber \\
 & & +z_1 \cdot x_2 \cdot z_3 x_4
+x_1 \cdot z_2 \cdot x_3 z_4 \nonumber \\
 & & -z_1 z_2 \cdot x_1 x_2 \cdot z_3 x_4 \cdot x_3 z_4,
\label{operator}
\end{eqnarray}
with eigenvalue nine.

We are now interested in the case in which the measurements are
imperfect and the observed correlation functions $E_{z_1 \cdot
z_3}$, $E_{z_2 \cdot z_4}$, \dots , $E_{z_1 z_2 \cdot x_1 x_2
\cdot z_3 x_4 \cdot x_3 z_4}$ fail to attain the values assumed in
the ideal case (i.e., $\left\langle {\psi }\right| z_1 \cdot z_3
\left|{\psi } \right\rangle = -1$, $\left\langle {\psi }\right|
z_2 \cdot z_4 \left|{\psi } \right\rangle = -1$, \dots ,
$\left\langle {\psi }\right| z_1 z_2 \cdot x_1 x_2 \cdot z_3 x_4
\cdot x_3 z_4 \left|{\psi } \right\rangle = -1$). We therefore
inquire whether the measured probability distribution functions
$P_{AB}(a,b)$ (with $A$ being the operator that Alice measures on
qubits 1 and 2, $B$ being the operator that Bob measures on qubits
3 and 4, and each $a$, $b$ being $-1$ or $+1$) that describe the
outcomes of the nine different experiments on the state
$\left|{\psi } \right\rangle$, can all be represented in the form
\begin{equation}
P_{AB}(a,b) = \int\!\rho (\lambda) p(a,\lambda) p(b,\lambda),
\label{locality}
\end{equation}
where $\lambda$ is a set of parameters common to the four
qubits, with distribution $\rho (\lambda)$, subject only to the
requirement that the outcome of an experiment performed by
Alice (Bob) for given $\lambda$
does not depend on Bob's (Alice's) choice of experiment.

If a representation (\ref{locality}) exists, then the mean of a
product of one of Alice's
measured operators, $A$, and one of Bob's, $B$,
will be given by
\begin{equation}
E_{AB} = \int\!\rho (\lambda) E_A(\lambda) E_B(\lambda),
\label{tf}
\end{equation}
where each $E$ in the integrand is of the form
\begin{equation}
E = p(+1,\lambda)-p(-1,\lambda).
\label{constraint}
\end{equation}
In particular, the linear combination of correlation functions
corresponding to the linear combinations of operators appearing in
the definition of $O$ [Eq.~(\ref{operator})] can be expressed as
\begin{eqnarray}
F = \int\!\rho (\lambda) (& & -E_{z_1 \cdot z_3} -E_{z_2 \cdot
z_4} -E_{x_1 \cdot x_3}
-E_{x_2 \cdot x_4} \nonumber \\
 & & +E_{z_1 z_2 \cdot z_3 \cdot z_4}
+E_{x_1 x_2 \cdot x_3 \cdot x_4} +E_{z_1 \cdot x_2 \cdot z_3 x_4}
\nonumber \\
 & & +E_{x_1 \cdot z_2 \cdot x_3 z_4}
-E_{z_1 z_2 \cdot x_1 x_2 \cdot z_3 x_4 \cdot x_3 z_4}).
\label{Floc}
\end{eqnarray}
According to quantum mechanics, in the state
$\left|{\psi } \right\rangle$, $F$ is given by
\begin{equation}
F_{\rm{QM}}=\left\langle {\psi }\right| O \left|{\psi } \right\rangle
= 9.
\label{qprediction}
\end{equation}
However, if it can be
expressed in the form (\ref{Floc})
there is a more restrictive bound on $F$.
Each of the 12 quantities
$E$ ($E_A$ or $E_B$) appearing in (\ref{Floc})
is constrained by
(\ref{constraint}) to lie between $-1$ and $+1$. Since the
integrand of (\ref{Floc}) is linear in each $E$ (keeping the other
11 fixed), it will take its extremal values when
the variables $E$ take their extremal values.
Therefore, as can be easily checked, if $F$
can be represented in the form (\ref{Floc}) then
\begin{equation}
F \le 7,
\label{EPRprediction}
\end{equation}
which contradicts the corresponding quantum prediction given by
(\ref{qprediction}).


The first eight experiments involved in the inequality consist of
local measurements of single spin components or single products of
two spin components on qubits prepared in the singlet state, and
do not entail any particular difficulty. To experimentally test
property (\ref{nueve}), or measure the corresponding correlation
function in the inequality (i.e., $E_{z_1 z_2 \cdot x_1 x_2 \cdot
z_3 x_4 \cdot x_3 z_4}$), it is not necessary to measure $z_1 z_2$
{\em and} $x_1 x_2$ on Alice's qubits, and $z_3 x_4$ {\em and}
$x_3 z_4$ on Bob's qubits. Each of such measurements is equivalent
\cite{Cabello01} to making a complete discrimination between four
Bell states in both spacelike separated regions. If the qubits are
polarized photons, such complete discrimination requires nonlinear
interactions \cite{VY99,LCS99}. An experiment of this kind has
been recently reported \cite{KKS01}. On the other hand, a setup
for performing joint measurements of $z_1 z_2$ {\em and} $x_1 x_2$
(or $z_3 x_4$ {\em and} $x_3 z_4$) for path and spin degrees of
freedom of a single particle has been proposed in \cite{SZWZ00}.
However, to experimentally test property (\ref{nueve}), it would
be enough to be able to measure the {\em product} of $z_1 z_2$ by
$x_1 x_2$ on Alice's qubits and the {\em product} of $z_3 x_4$ by
$x_3 z_4$ on Bob's qubits. These measurements are, respectively,
equivalent to measuring $y_1 y_2$ and $y_3 y_4$ (being
$y_i=\sigma_{yi}$), and could therefore be performed locally by
measuring $y_1$ and $y_2$ and multiplying their results, and by
measuring $y_3$ and $y_4$ and multiplying their results. As can
easily be checked, results $y_1 y_2=\pm 1$ are respectively
equivalent to results $z_1 z_2 \cdot x_1 x_2=\mp 1$, and results
$y_3 y_4=\pm 1$ are respectively equivalent to results $z_3 x_4
\cdot x_3 z_4=\pm 1$. However, while it is not difficult to
perform spin measurements along either $x$ or $z$ directions on a
qubit flying along direction $y$, spin measurements along $y$ face
several problems. A different solution arises from the observation
that distinguishing between the results $+1$ and $-1$ when
measuring $z_1 z_2 \cdot x_1 x_2$ is equivalent to distinguishing
between, respectively, the {\em pairs} of Bell states $\left\{
\left| {\phi ^+} \right\rangle _{12}, \left| {\psi ^-}
\right\rangle _{12} \right\}$ and $\left\{ \left| {\phi ^-}
\right\rangle _{12}, \left| {\psi ^+} \right\rangle _{12}
\right\}$, where
\begin{eqnarray}
\left| \phi^\pm \right\rangle & = &
{1 \over \sqrt{2}}
\left( {
\left| 00 \right\rangle \pm
\left| 11 \right\rangle
} \right),
\label{Phi} \\
\left| \psi^\pm \right\rangle & = &
{1 \over \sqrt{2}}
\left( {
\left| 01 \right\rangle \pm
\left| 10 \right\rangle
} \right).
\label{Psi}
\end{eqnarray}
Analogously,
distinguishing between the results $+1$ and $-1$
when measuring
$z_3 x_4 \cdot x_3 z_4$ is
equivalent to distinguishing between,
respectively, the
{\em pairs} of Bell states
$\left\{ \left| {\chi ^+} \right\rangle _{34},
\left| {\omega ^-} \right\rangle _{34} \right\}$ and
$\left\{ \left| {\chi ^-} \right\rangle _{34},
\left| {\omega ^+} \right\rangle _{34} \right\}$,
where
\begin{eqnarray}
\left| \chi^\pm \right\rangle & = &
{1 \over \sqrt{2}}
\left( {
\left| 0\bar{0} \right\rangle \pm
\left| 1\bar{1} \right\rangle
} \right),
\label{Chi} \\
\left| \omega^\pm \right\rangle & = &
{1 \over \sqrt{2}}
\left( {
\left| 1\bar{0} \right\rangle \pm
\left| 0\bar{1} \right\rangle
} \right),
\label{Omega}
\end{eqnarray}
where $x \left| \bar{0} \right\rangle =
\left| \bar{0} \right\rangle$ and
$x \left| \bar{1} \right\rangle =
-\left| \bar{1} \right\rangle$.
Therefore, previous setups
involving only linear elements
which distinguish two out four Bell
states for photons entangled in polarization
\cite{VY99,LCS99,dc2,telep2} could be used
to test property (\ref{nueve}).


Returning to the gedanken version, the fact that similar proofs of
Bell's theorem can be developed for every common eigenstate of
$z_1 z_3$, $z_2 z_4$, $x_1 x_3$, and $x_2 x_4$, leads us to wonder
whether our gedanken proof of Bell's theorem could be the basis
for a state-independent proof of the KS theorem on the
impossibility of ascribing noncontextual hidden variables (i.e.,
those which assign predefined values to the physical observables,
assuming that such values do not depend on which other compatible
observables are jointly measured) to quantum mechanics
\cite{KS67}. Mermin has derived such proofs of the KS theorem both
from a previous state-dependent proof of the KS theorem by Peres
\cite{Peres90} and from his own simplification
\cite{MerminGHZa,MerminGHZb} of the GHZ proof
\cite{Mermin90a,Mermin93}.

Table~I contains a state-independent proof of the KS theorem based
on the ``all versus nothing'' proof of Bell's theorem for two
observers introduced in this paper. The main difference between
Mermin's proof of the KS theorem inspired by Peres' \cite{Peres90}
and the proof in Table~I is that, while the former has two rows
containing nonlocal operators, in the latter only the first row
contains nonlocal operators (i.e., those which cannot be measured
by only one observer). This implies that, while the former cannot
be transformed into proof of Bell's theorem, the latter (and the
one derived by Mermin from the GHZ proof) can be converted into
proof of Bell's theorem.


In brief, I have shown that the Hardy-like proof presented in
Ref.~\cite{Cabello01} can be rearranged as a GHZ-like proof with
only two observers which, on one hand, allows us to derive an
experimentally testable Bell inequality and, on the other hand,
leads to a new state-independent proof of the KS theorem. Thus
Ref.~\cite{Cabello01} and this paper provide a wider perspective
on the relations between the major no-hidden-variables theorems
(KS's and Bell's) and their proofs (KS's state-independent
\cite{KS67}, Bell's with inequalities \cite{Bell64,CHSH69},
Hardy's without inequalities but with probabilities
\cite{Hardy93}, and GHZ's without inequalities or probabilities
\cite{GHZ89,MerminGHZa,MerminGHZb}).


This work was sparked by a question proposed by N. Gisin.
I would like to thank J. Calsamiglia, J. L. Cereceda,
N. D. Mermin, and A. Peres for useful comments.
This work was supported in part by
the Spanish Ministerio de Ciencia y Tecnolog\'{\i}a
(Grant No.\ BFM2000-0529).



\begin{table}
\begin{center}
\begin{tabular}{ccccccccc}
\hline
\hline
$z_1 z_3$ & & $z_2 z_4$ & & $x_1 x_3$ & & $x_2 x_4$ & & $y_1 y_2 y_3 y_4$ \\
$z_1$ & & $z_2$ & & & & & & $z_1 z_2$ \\
 & & & & $x_1$ & & $x_2$ & & $x_1 x_2$ \\
$z_3$ & & & & & & $x_4$ & & $z_3 x_4$ \\
 & & $z_4$ & & $x_3$ & & & & $x_3 z_4$ \\
\hline
\hline
\end{tabular}
\end{center}
\noindent TABLE I.
{\small Proof of the Kochen-Specker theorem.
Each row or column contains mutually commutative operators.
The product of the operators of each row or column is the
identity except for the last column which is minus the identity.
We cannot assign noncontextual values, either $-1$ or $+1$, to
each of the 17 operators appearing in the table
if we assume that the product of
these values must be $+1$ for all the rows and columns except for the
last column which must be $-1$.
This is so because, when the product of
the values appearing in all the rows and columns is taken, each value appears
twice while the product of all of them ought to be $-1$.}
\end{table}

\end{document}